\documentstyle[preprint,aps]{revtex}
\begin{document}
\draft
\title{Elastic Co--Tunneling in a 1D Quantum Hole Approach}
\author{Heinz-Olaf M{\"u}ller, Andreas H{\"a}dicke, and Wolfram Krech}
\address{Institut fr Festk{\"o}rperphysik\\
        Friedrich--Schiller--Universit{\"a}t Jena\\
        Max--Wien--Platz 1\\
        07743 Jena}
\date{\today}
\maketitle
\begin{abstract}
The influence of the phases of tunneling matrix elements on the rate of
the elastic co--tunneling at an ultrasmall normal--conducting
double--junction is studied in a simple quantum--hole approach at zero
temperature. The results are compared with experimental data.
\end{abstract}

It is known that besides ordinary tunneling of single charges through
ultrasmall double--junctions there is the effect of macroscopic quantum
tunneling (q--MQT) via virtual intermediate states~\cite{ave7},~\cite{ave8}.
Though the current of q--MQT is much smaller than that of ordinary
tunneling, it can be observed for voltages below the Coulomb--blockade,
where ordinary tunneling is impossible~\cite{han1}. Here we consider only
the so--called elastic channel of q--MQT that is qualified by a nonvanishing
conductivity at zero voltage as predicted theoretically.
Averin and Nazarov~\cite{ave7},~\cite{ave8} have presented a
theory of elastic q--MQT basing on~\cite{gor1}
that takes into consideration the phases of tunnel
matrix elements yielding the following expression for the conductivity
$G_{el}$ at low voltages $V$ and temperatures $T$
\begin{equation}\label{averin}
G_{el} = \frac{\hbar
G_1G_2\Delta}{4\pi\,e^2}\left(\frac{1}{E_1}+\frac{1}{E_2}\right).
\end{equation}

In this paper we try to give an modified theory of elastic q--MQT where
the central electrode is approximated by an 1D quantum hole. This is
justified because of its size. We study a simple double--junction,
consisting of two ultrasmall tunnel junctions connected in series and
possessing capacities $C_1$, $C_2$ ($C_{\Sigma}=C_1+C_2$), driven by the
voltage $V$. It is assumed that the external electrodes are metallic bulks.
Finally, the result is a formula for the conductivity including the discrete
island level spacing $\Delta_0$ at the Fermi edge. Note, that in the continuous
case $\Delta_0^{-1}$ corresponds to the energy density of states
at the Fermi level. $G_{1,2}$ denote the individual tunnel conductivities of
the corresponding junctions. $E_1$ and $E_2$ are the  electrostatic energy
differences connected with tunneling at the respective electrodes and read
as
\begin{equation}
E_{1,2} = \pm\frac{e^2}{C_{\Sigma}}\left[n\pm\frac{1}{2}
\mp\frac{C_{2,1}V}{e}\right].
\end{equation}
$n\,e$ is the integer--valued charge on the central electrode due to
tunneling. In terms of a first order Taylor expansion in the supplied
voltage $V$ the elastic conductivity $G_{el}$ turns out to
be at low temperatures~\cite{ave7},~\cite{ave8}
\begin{eqnarray}
G_{el} & = & \frac{2\pi e^2}{\hbar}\sum_{k\,k'}T_{k\,k'}^{(1)}
T_{k'\,0}^{(1)\ast}T_{0\,k}^{(2)\ast}T_{0\,k'}^{(2)}D_0^2
F(\varepsilon_k)F(\varepsilon_{k'}),\label{gel}\\
F(\varepsilon) & = & \frac{1-f(\varepsilon)}{E_1+\varepsilon}-
\frac{f(\varepsilon)}{E_2-\varepsilon}.\nonumber
\end{eqnarray}
Here $f$ denotes the Fermi function and $D$ the bulk electron density of
states. The subscript "0" indicates the Fermi level. In the coordinate
representation the tunneling amplitudes $T$ are expressed by the wave
functions $\psi$ in the following way
\begin{equation}
T_{km} = \int\mbox{d}^3y\mbox{d}^3z\,T(y,z)\psi^{\ast}_k(y)\psi_m(z).
\end{equation}
Note, that the subscripts $k$ and $l$ and the coordinates $y$ belong to the
central electrode of the double--junction, whereas the subscripts $m$, $n$
and the coordinates $z$ are used to describe the banks.

Following~\cite{ave7},~\cite{ave8} the conductivities of
the single--junctions in Golden--rule type estimation are given by
\begin{equation}
G_{1,2} = \frac{4\pi e^2}{\hbar}|T^{(1,2)}|^2\frac{D_0}{\Delta_0},
\end{equation}
where $D_0$ is the density of states of the banks at Fermi level. In
coordinate representation this looks like
\begin{equation}\label{g12}
G_1 = \frac{4\pi e^2}{\hbar}\int\mbox{d}^3y_{1,2}\mbox{d}^3z_{1,2}
T^{(1)}(y_1,z_1)T^{(1)\ast}(y_2,z_2)\psi_0^{\ast}(y_1)\psi_0(z_1)
\psi_0^{\ast}(z_2)\psi_0(y_2)\frac{D_0}{\Delta_0},
\end{equation}
and respectively in the case of the other junction.

Owing to the fact that the tunneling matrix elements $T(y,z)$ drop
exponentially outside of the vicinity of the junctions~\cite{ave8}, the
integrations with respect to $y$ contribute only for fixed values
\begin{equation}
y_{1,2} = -\frac{L}{2}\mbox{\hspace{5cm}}y_{3,4} = \frac{L}{2}.
\end{equation}
Then, $G_{el}$ simplifies to
\begin{equation}
G_{el} = \frac{\hbar}{8\pi\,e^2}G_1G_2\frac{
\sum_{k\,k'}\psi_k^{\ast}(-L/2)\psi_k(L/2)\psi_{k'}^{\ast}(L/2)\psi_{k'}
(-L/2)F(\varepsilon_k)F(\varepsilon_{k'})\Delta_0^2}
{\psi_0^{\ast}(-L/2)\psi_0(-L/2)\psi_0^{\ast}(L/2)\psi_0(L/2)}.
\end{equation}
Using real wave functions according to~\cite{ave8} $G_{el}$ may be factorized
in this approximation with regard to $k$ and $k'$, resulting in
\begin{equation}\label{gel1}
G_{el}=\frac{\hbar}{8\pi e^2}G_1G_2\left[
\frac{\sum_k\psi_k(-L/2)\psi_k(L/2)F(\varepsilon_k)}{\psi_0(-L/2)\psi_0(L/2)}
\Delta_0\right]^2.
\end{equation}

Now, the estimation of (\ref{gel1}) is done within a quantum hole
approximation. This is different to the original approach of Averin and
Nazarov. The central electrode is taken as 1D quantum hole with the
length $L$ and the energetic depth $E$ (counted from the Fermi level,
$E>0$).

Now, the solution of the quantum mechanical problem is done within the
variables~\cite{flu1}
\begin{equation}
E=\frac{\hbar^2k_E^2}{2\,m^{\ast}},\hspace{1cm}
\kappa^2=k_E^2-k^2.
\end{equation}
In these terms the wave functions are
\begin{eqnarray}
\psi_+(x) & = & \left\{\begin{array}{ll}
A_+\cos k\,x & 0\le|x|\le L/2\\
A_+\cos\frac{k\,L}{2}\mbox{e}^{\kappa(L/2-|x|)} & |x|>L/2
\end{array}\right.\nonumber\\
\psi_+(-x) & = & \psi_+(x)\nonumber\\
\frac{1}{A_+^2} & = & \frac{1}{k}\left[\frac{k\,L}{2}+\sin\frac{k\,L}{2}\cos
\frac{k\,L}{2}\right]+\frac{1}{\kappa}\cos^2\frac{k\,L}{2}\nonumber\\
\psi_-(x) & = & \left\{\begin{array}{ll}
A_-\sin k\,x & 0\le|x|\le L/2\\
A_-\sin\frac{k\,L}{2}\mbox{e}^{\kappa(L/2-|x|)} & |x|>L/2
\end{array}\right.\nonumber\\
\psi_-(-x) & = & -\psi_-(x)\nonumber\\
\frac{1}{A_-^2} & = & \frac{1}{k}\left[\frac{k\,L}{2}-\sin\frac{k\,L}{2}\cos
\frac{k\,L}{2}\right]+\frac{1}{\kappa}\sin^2\frac{k\,L}{2}.\nonumber
\end{eqnarray}
The finite discrete energy spectrum of the quantum hole results from the
solutions of the transcendental equations
\begin{eqnarray}\label{cond1}
\tan\frac{k_+L}{2} & = & \frac{\kappa}{k_+},\\
\label{cond2}
\tan\frac{k_-L}{2} & = & -\frac{k_-}{\kappa}
\end{eqnarray}
for symmetric ($+$) and antisymmetric ($-$) wave functions, respectively.
Due to the low conductivities of the single junctions the wave function of
the island is well approximated by the wave function of the quantum hole.

According to~\cite{sim1}, ohmic behaviour of the current can be expected in
case of $e\,V\ll E$, only. Hence, we simplify the model in terms of
$k\ll\kappa$ which corresponds to $e\,V\ll E$ via an uncertainty relation.
In this approximation the solutions of Eq. (\ref{cond1})  and (\ref{cond2})
read as
\begin{equation}\label{kappr}
k_n\approx\frac{\pi\,n}{L}\left(1-\frac{2}{\kappa\,L}\right),\quad
n=0,1,2,\ldots n_E
\end{equation}
with
$\hbar\kappa\approx 2\,m^{\ast}\sqrt{2\,m^{\ast}(E-\varepsilon_F)}/\hbar$.
$n_E$ denotes the total number of states in the quantum hole of the
energetic depth $E$. Furthermore, the following Taylor expansion
in $1/(\kappa L)$ of the required wave function values can be obtained~:
\begin{eqnarray}\label{last}
\psi_{\pm}(L/2)\psi_{\pm}(-L/2) & = & \pm A_{\pm}^2\left\{
\begin{array}{c} \cos^2(k\,L/2)\\ \sin^2(k\,L/2)
\end{array}\right\}\\
& \approx & \pm
\left(\frac{\pi\,n}{\kappa\,L}\right)^2\left[1-\frac{2}{\kappa\,L}
-\frac{1}{3}\left(\frac{\pi\,n}{\kappa\,L}\right)^2\right]+{\cal O}
\left(\frac{1}{(\kappa L)^6}\right).\nonumber
\end{eqnarray}
Fig.~\ref{fig0} shows a comparison of different order contributions to
$\psi(L/2)\psi(-L/2)$ at $\varepsilon(k_n) = \varepsilon_F$.

By means of (\ref{last}) and the approximation (\ref{kappr})
Eq.~(\ref{gel1})
is given for zero temperature by
\begin{eqnarray}\label{gel3}
G_{el} & = & \frac{\hbar}{8\pi e^2}G_1G_2\left[\sum_{n=1}^{n_E}
(-1)^n\frac{E-\varepsilon_F}{E-\varepsilon_F-\varepsilon_n}
\Delta_0F(\varepsilon_n)\right]^2\\
& = & \frac{\hbar}{8\pi e^2}G_1G_2\left[
\sum_{n=1}^{n_F}
(-1)^n\frac{E-\varepsilon_F}{E-\varepsilon_F-\varepsilon_n}\,
\frac{\Delta_0}{\varepsilon_n-E_2}+\right.\nonumber\\
& & \left.+\sum_{n=n_F+1}^{n_E}
(-1)^n\frac{E-\varepsilon_F}{E-\varepsilon_F-\varepsilon_n}\,
\frac{\Delta_0}{\varepsilon_n+E_1}
\right]^2.\nonumber
\end{eqnarray}
Here, $n_F$ labels the number of states below the Fermi level ($n_F\le
n_E$). The contributions of neighboring states possess different signs due
to the different symmetry of the respective wave functions. This effect is
known from~\cite{ave8}. Formula (\ref{gel3}) is our main result. It is much
more complicated than expression (\ref{averin}) because the discrete energy
spectrum of the central electrode (quantum hole) has been taken into account.

In discussion we are going to compare our results with the experimental data
of~\cite{han1}. For the evaluation numerical values of $n_F$ and
$\varepsilon_F$ are necessary. According to the experimental setup, the values
are chosen as $n_F =  5.9\times10^{22}cm^{-3}$ and $\varepsilon_F =
5.51 eV$~\cite{kit1}. The quasi--particle mass $m^{\ast}$ is approximated by
the free electron mass $m^{\ast}\approx m_0 \approx 9.1\times10^{-31}kg$.
The parameter $L$ corresponding to the diameter of the central electrodes is
given by the experiment~\cite{han1}. The second parameter $E$ describing the
barrier height of the quantum hole has to be determined in comparison with
the experimental data. Fig.~\ref{fig1} and Fig.~\ref{fig2} show the computed
elastic conductivity $G_{el}$ of sample $A$ and $B$ of~\cite{han1} in
dependence on $E$, respectively. The observed steps arise from the
occupation of a new state in the quantum hole. The parameter $E$ is given by
the figures where the computed $E$--dependent conductivity $G_{el}$
coincides with the experimental value (independent of $E$). Due to the
fluctuations of the numerical curves a least--squares fit was applied to
them. This yields the corresponding parameters $E$ of the samples $A$ and
$B$; $E_A \approx E_B \approx 2\ldots3\,eV$ (cf.\ Fig.~\ref{fig1} and
Fig.~\ref{fig2}). This is in good agreement with earlier
estimations~\cite{gun1}.

The pros of this model are the simple theory behind it. In addition
to~\cite{ave8} our result depends on properties of the island
characterized by $E$ and $L$.
According to the applied model, $\Delta_0 \approx 0.38eV$, $1.12eV$ is found
for sample $A$ and $B$, respectively. These values exceed drastically the
data concluded from the experiment~\cite{han1} in terms of the
theory~\cite{ave8} ($\Delta=0.045eV$, $0.126eV$). This is not surprising
because they are model--dependent parameters.

Our approach is based on the very simple model of an
1D quantum hole, which is far from being general due to the neglect of two
dimensions and the idealization of the energetic shape of the barriers.
Regarding to the second point it is unlikely within $eV\ll E$ that the shape
of the upper end of barriers influences the current dramatically. An
influence on the steps observed in Fig.~\ref{fig1} and Fig.~\ref{fig2}
could be expected.
This approach could be improved by using more sophisticated models of the
barrier, as presented for instance by~\cite{gun1},~\cite{kna1}.
Additionally, instead of the bulk values for $\varepsilon_F$ and $n_F$ as
given above more appropriated values should be taken from the experiment.

This work was supported by the Deutsche Forschungsgemeinschaft.

\begin{figure}
\caption{The evaluation of the different orders of a Taylor expansion of
the wave function in a deep quantum hole. -- The solid line corresponds to
the numerical solution whereas the other curves display the Taylor
expansion of given order.}
\label{fig0}
\end{figure}

\begin{figure}
\caption{The comparison of our formulae with the experimental data of
sample $A$. -- The figure shows the dependence of the elastic conductivity
$G_{el}$ on the barrier height $E$. We compare the numerical solution
(solid line) with the fitted value out of the experimental paper
(horizontal line). The steps indicate the occupation of a new state within
the quantum hole. The straight line corresponds to a $G_{el} = a\,E^{b}$
least--square fit of the numerical data. The parameters of the sample are $L
= 3\,nm$, $C_1 = 0.57\times10^{-18}F$, $C_2 = 3.4\times10^{-18}F$, $G_1 =
1/(3.5 M\Omega)$, and $G_2 = 1/(21 M\Omega)$.}
\label{fig1}
\end{figure}

\begin{figure}
\caption{The corresponding comparison of data like Fig.~1 for
sample $B$. Its parameters are $L = 0.8\,nm$, $C_1 = 0.095\times10^{-18}F$,
$C_2 = 0.066\times10^{-18}F$, $G_1 = 1/(0.42 M\Omega)$, and $G_2 =
1/(630 M\Omega)$.}
\hfill\label{fig2}
\end{figure}


\begin{references}
\bibitem{ave7}
D.~V. Averin and Y.~V. Nazarov, Phys. Rev. Lett. 65 (1990) 2446.

\bibitem{ave8}
D.~V. Averin and Y.~V. Nazarov, in~: Single Charge Tunneling: Coulomb
  Blockade Phenomena in Nanostructures (NATO ASI Series B: Physics, Vol.
  294), eds. H. Grabert, M.~H. Devoret (Plenum Press, New York and
  London, 1992), pp.\ 217--247.

\bibitem{han1}
A.~E. Hanna, M.~T. Tuominen, and M. Tinkham, Phys. Rev. Lett. 68 (1992) 3228.

\bibitem{gor1}
L.~P. Gorkov and G.~M. Eliashberg, Zh. Eksp. Teor. Fiz. 48 (1965) 1402.

\bibitem{flu1}
S. Fl{\"u}gge, Practical Quantum Mechanics, Vol.~1 (Springer, Berlin,
  Heidelberg, 1971).

\bibitem{sim1}
J.~G. Simmons, J. Appl. Phys. 34 (1963) 2581.

\bibitem{kit1}
C. Kittel, Einf{\"u}hrung in die Festk{\"o}rperphysik (Oldenbourg,
  M{\"u}nchen, Wien, 1989).

\bibitem{gun1}
K.~H. Gundlach and J. H{\"o}lzl, Surf. Science 27 (1971) 125.

\bibitem{kna1}
H. Knauer, J. Richter, and P. Seidel, Theorie analytischer Strombeziehungen
fr kleine Spannungen und Anwendung zur Bestimmung charakteristischer
Barrierenparameter von MIM--Tunnelstrukturen, unpublished (1978).
\end{references}
\end{document}